\documentstyle[12pt]{article}
\newcommand{\La}{\Lambda}
\newcommand{\PL}{P_{\Lambda}}
\newcommand{\PR}{P_{R}}
\newcommand{\PiR}{\Pi_{R}}
\newcommand{\PibR}{\Pi_{\bar R}}
\newcommand{\bR}{\bar R}
\begin{document}
\baselineskip=22pt plus 0.2pt minus 0.2pt
\lineskip=22pt plus 0.2pt minus 0.2pt
\font\bigbf=cmbx10 scaled\magstep3
\begin{center}
 {\bigbf Kruskal coordinates as canonical variables for Schwarzschild black 
holes}

\vspace*{0.35in}

\large

Madhavan Varadarajan
\vspace*{0.25in}

\normalsize

{\sl Raman Research Institute,
Bangalore 560 080, India.}
\\
madhavan@rri.ernet.in\\
\vspace{.5in}
November 2000\\
\vspace{.5in}
ABSTRACT

\end{center}
We derive a transformation from the usual ADM metric-extrinsic
curvature variables on the phase space of Schwarzschild black holes, to new
canonical variables which  have the interpretation of 
Kruskal coordinates. We explicitly show that this transformation 
is non-singular, even at the horizon. 
The constraints of the theory simplify in terms of the new 
canonical variables and are equivalent to the vanishing of the canonical 
momenta. 

Our work is based on earlier seminal work by Kucha{\v r} in which he 
reconstructed curvature coordinates and a mass function from spherically
symmetric canonical data. The key feature in our construction of a nonsingular
canonical transformation to Kruskal variables, is the scaling of the curvature 
coordinate variables
 by the mass function rather than by the mass at left spatial 
infinity.

\pagebreak

\setcounter{page}{1}

\section*{1. Introduction} 
This work is devoted to an extension and improvement of Kucha{\v r}'s 
beautiful analysis of the Hamiltonian description of Schwarzschild 
black holes. In \cite{kuchar}  Kucha{\v r} reconstructed the curvature 
coordinates (i.e. the areal radius and the Killing time)  as well as a
mass function from spherically symmetric ADM canonical data on a Cauchy slice. 
The curvature coordinates were turned into canonical variables
and the constraints in the Hamiltonian description simplified when expressed
in terms of the new canonical variables. Since the curvature coordinates
are not good spacetime coordinates on the horizon, Kucha{\v r}'s canonical 
transformation is singular on the horizon. Nevertheless, it was argued in 
\cite{kuchar} that, with sufficient care near the horizon, the curvature 
coordinate variables 
could be used to simplify the Hamiltonian description and that 
the imposition of the constraints was equivalent to the vanishing of the 
 momenta conjugate to the curvature coordinate variables. 

In this work we improve upon the treatment of \cite{kuchar} by constructing a 
 transformation to new canonical variables which have the interpretation of
Kruskal coordinates. This transformation is free from the singularities 
of the canonical transformation to curvature coordinate variables.
\footnote{ 
In \cite{kuchar}, a canonical transformation on the curvature 
coordinate variables involving scaling these variables by the mass at left 
spatial infinity yielded new canonical variables which, on shell, also
had the interpretation of Kruskal coordinates. However, as discussed in
section 7, the transformation from ADM variables to these Kruskal variables
is still not free from singularities. In contrast, our canonical transformation
involves scaling of the curvature coordinate variables by the 
{\em mass function}. It is this feature which enables a nonsingular
description.
}
The constraints simplify in terms of the new Kruskal variables and  their 
imposition is equivalent to the vanishing of the new canonical momenta.
As in \cite{kuchar}, the true degrees of freedom are the mass at left infinity
 and the difference between Killing time and parametrization time at 
left infinity.

The layout of the paper is as follows. In section 2 
we quickly review the relevant parts of \cite{kuchar}. 
The purpose of this review is to establish  notation and to collect the set of 
equations from \cite{kuchar} which we shall use to establish our results.
The reader may consult \cite{kuchar} for more details and we shall assume
familiarity with that work.
In section 3 we derive 
the 
canonical transformation to the Kruskal variables  from the curvature 
coordinate variables of \cite{kuchar} and  express the constraints 
in terms of the Kruskal variables. 
In section 4, we express the ADM variables in terms of the new canonical 
variables and note that the transformation is manifestly non singular at the
horizon. 

We describe our choice of asymptotic behaviour for the canonical 
variables
 in section 5. In section 6 we invert the transformation 
of section 4 and express the Kruskal variables in terms of the ADM variables.
Section 7 contains our 
concluding remarks.

\section*{2. Review of Kucha{\v r}'s results.}
In this section we briefly review the results of \cite{kuchar}. As 
mentioned in the introduction,
the purpose of this section is to establish notation and collect  
the set of equations from \cite{kuchar} which  we shall use to establish our 
results. The reader may consult \cite{kuchar} for more details.

Spherically symmetric Cauchy slices 
in the Hamiltonian description of spherically symmetric gravity
are diffeomorphic to $S^2 \times {\bf R}$. The spatial metric induced on 
such a slice  is 
\begin{equation}
d\sigma^2 =\Lambda^2(r) (dr)^2 + R^2(r) (d\Omega )^2
\label{metric}
\end{equation}
where $r$ is a radial coordinate and $d\Omega$ is the line element on 
the unit sphere. $r=\infty$ labels right spatial infinity and $r=-\infty$
labels left spatial infinity.
$\PL (r) $ and $\PR (r)$ are the momenta conjugate to $\Lambda (r)$ and 
$R (r)$. After parametrizing the times at the two spatial infinities, the 
action takes the form
\footnote{
We denote derivatives with respect to `$t$' by a dot
and spatial derivatives with respect to `$r$' by a prime.
We use units in which Newton's constant, $G$, the speed of light, $c$ and
Planck's constant, $\hbar$, are unity.
}
\begin{eqnarray}
S(\La,\PL, R,\PR, N, N^r, \tau_+, \tau_-) 
&=& \int dt \int_{-\infty}^{\infty} dr
    (\PL {\dot{\La}} + \PR {\dot{R}} -N H - N^rH_r)\nonumber \\ 
&-& \int dt (M_+ {\dot{\tau_+}}-M_- {\dot{\tau_-}}).
\label{1}
\end{eqnarray}
$H_r$ is the radial diffeomorphism constraint and  $H$ is 
the scalar constraint. They are 
given by 
\begin{equation}
H_r = \PR R^{\prime} -\La \PL ^{\prime} 
\label{2}
\end{equation}
and
\begin{equation}
H= -R^{-1} \PR \PL + {1\over 2} R^{-2 }\La \PL^2
  + \La^{-1} R R^{\prime\prime}- \La^{-2}\La^{\prime} RR^{\prime} 
  + {1\over 2} \La^{-1}R^{\prime 2} - {1\over 2} \La .
\label{3}
\end{equation}
The parameters $\tau_+$ and $\tau_-$ label parametrization clocks at 
right and left spatial infinity and are to be freely varied in the action
as are the lapse function, $N$, and the radial shift vector field, $N^r$. 
$M_{\pm}$ are parameters
which appear in the asymptotic fall off of $\La$ at right and left spatial 
infinity.

To leading order in $r$, the asymptotic behaviour of the canonical 
coordinates as $r\rightarrow \pm\infty$ on any Cauchy slice of constant `$t$'
is
\begin{equation}
\La = 1 + M_{\pm} |r|^{-1}, \;\;\; R =|r| .
\end{equation}

By parametrizing the standard form of the Schwarzschild line element in 
curvature coordinates in terms of arbitrary parameters $(t,r)$, comparing 
the result with the standard ADM form of the line element and using the 
relation between canonical momenta and the velocities of canonical coordinates,
Kucha{\v r} was able to reconstruct the Schwarzschild mass and the 
rate of change of Killing time from the canonical data. The 
expression for the Schwarzschild mass function is 
\begin{equation}
M = {1\over 2} R^{-1} \PL^2 - {1\over 2} \La^{-2}RR^{\prime 2} +
{1\over 2} R .
\label{4}
\end{equation}
It is useful to define the quantity `$F$' by 
\begin{eqnarray}
F &=& 1 - {2M\over R}
\label{5} \\
&=& ({R^\prime\over \La})^2 - ({\PL\over R})^2 .
\label{6}
\end{eqnarray}
When the constraints are satisfied, the vanishing of $F$ determines the 
location of horizons of the (extended) Schwarzschild spacetime. We shall 
continue to use the word `horizon' when $F=0$ even when the constraints are 
not imposed.

Denoting the Killing time by $T$, its rate of change along the slice is
given by 
\begin{equation}
T^{\prime} = -R^{-1} F^{-1} \La \PL .
\label{7}
\end{equation}
$T$ is turned into a canonical coordinate by making a transformation 
from \\
\noindent $(\La (r), \PL(r), R(r), \PR(r), \tau_{\pm})$ to the variables
$(T(r), P_T(R) , R(r),\PiR (r),m, p)$ 
\footnote{
Our notation differs from the notation of  \cite{kuchar} in that
${\rm R},P_{\rm R}$ of \cite{kuchar} are denoted by $R, \PiR$ in this work.
}  
\\
\noindent where
\begin{equation}
T(r) =\tau_+ + \int_{\infty}^{r} T^{\prime}({\bar{r}}) d {\bar{r}}  ,
\label{8}
\end{equation}
\begin{eqnarray}
P_T(r) &=& - M^{\prime}(r) ,
\label{9}\\
&=& \La^{-1} (R^{\prime} H + R^{-1}\PL H_r), \label{9a}
\end{eqnarray}
\begin{eqnarray}
\PiR &= &P_R - {1\over 2} R^{-1} \La \PL - {1\over 2}R^{-1} F^{-1}\La \PL
\nonumber \\       
&-& R^{-1} \La^{-2} F^{-1} ((\La \PL)^\prime (RR^{\prime})  
                                 - \La \PL (RR^{\prime})^\prime)
\label{10} \\
&=& F^{-1}(R^{-1}\PL H +R^{\prime}\La^{-2} H_r) ,
\label{11}
\end{eqnarray}
\begin{equation}
p= T(-\infty ) -\tau_- ,
\label{11a}
\end{equation}
\begin{equation}
m= M_- .
\label{11b}
\end{equation}
The action in (\ref{1}) is replaced by 
\begin{eqnarray}
S(T, P_T, R, \PiR, N, N^r, m, p) 
&=& \int dt(p{\dot{m}} +
 \int_{-\infty}^{\infty} dr
    (P_T {\dot{T}} + \PiR {\dot{R}})) \nonumber \\
&-& \int dt \int_{-\infty}^{\infty} dr(N H - N^rH_r) .
\label{11c}
\end{eqnarray}
The constraints take the form
\begin{equation}
H_r= \PiR R^{\prime} +P_T T^{\prime},
\label{12}
\end{equation}
\begin{equation}
\La H = F^{-1}P_TR^{\prime} + FT^{\prime}\PiR,
\label{13}
\end{equation}
with 
\begin{equation}
\La = (F^{-1} R^{\prime 2} -FT^{\prime 2})^{1\over 2}.
\label{14}
\end{equation}
From (\ref{7}),(\ref{11}) and (\ref{14}), it can be seen that 
when $F\rightarrow 0$, the transformation becomes singular.

\section*{3. Canonical transformation to Kruskal variables.}
In this section we construct a canonical tranformation from the curvature 
coordinate variables $(R, \PiR, T, P_T)$ to Kruskal variables 
$( U, P_U, V, P_V)$.  The transformation is performed 
in two steps. The curvature 
coordinate variables are scaled with the mass function, $M(r)$, in the
first step and in the second, a point transformation is made from the 
scaled variables to the Kruskal variables.

In what follows, we shall make use of the identity
\footnote{
For a proof of the identity, see \cite{kuchar}
}:
\begin{equation}
\int_{-\infty}^{\infty} dr f(r)\int_{-\infty}^{r}d{\bar r}g({\bar r}) 
=-\int_{-\infty}^{\infty} dr g(r)\int_{\infty}^{r}d{\bar r}f({\bar r}) .
\label{identity}
\end{equation}

The Liouville form, $\omega$, in terms of the curvature coordinate 
variables is 
\begin{equation}
\omega = p{\dot{m}} +
 \int_{-\infty}^{\infty} dr
    (P_T {\dot{T}} + \PiR {\dot{R}}).
\end{equation}
Using (\ref{identity})
and (\ref{9}), and ignoring total time derivatives, it can be shown  that
\begin{eqnarray}
\int_{-\infty}^{\infty} dr\PiR {\dot{R}}
&=&\int_{-\infty}^{\infty} dr(2M\PiR){d\over dt}({R\over 2M})
+\int_{-\infty}^{\infty} dr{\PiR R\over M^2}m{\dot{m}} \nonumber\\
&&- \int_{-\infty}^{\infty} dr(2MP_T) 
            {d\over dt}(\int_{\infty}^rd{\bar r}{\PiR R\over 2M^2}),
\label{prr}
\end{eqnarray}
and that 
\begin{eqnarray}
\int_{-\infty}^{\infty} dr P_T {\dot{T}}
&=& \int_{-\infty}^{\infty} dr(2MP_T){d\over dt}({T\over 2M})
+\int_{-\infty}^{\infty} dr{P_T T\over M^2}m{\dot{m}} \nonumber \\
&&- \int_{-\infty}^{\infty} dr(2MP_T) 
            {d\over dt}(\int_{\infty}^rd{\bar r}{P_T T\over 2M^2}).
\label{ptt}
\end{eqnarray}
Equations (\ref{prr}) and (\ref{ptt}) imply that, upto a total time 
derivative,
\begin{equation}
\omega = {\bar p}{\dot{m}} +
 \int_{-\infty}^{\infty} dr
    (P_{\bar T} {\dot{\bar T}} + \PibR {\dot{\bar R}}),
\end{equation}
where the new canonical variables 
are given by 
\begin{eqnarray}
{\bar p}& =& p +m \int_{-\infty}^{\infty}dr({P_T T +\PiR R\over M^2}),
\label{16}\\
{\bar R}&=& {R\over 2M},
\label{17}\\
\PibR &=& 2M\PiR,
\label{18}\\
{\bar T}& = & {T\over 2M} - 
\int_{\infty}^{r}d{\bar r}({P_T T +\PiR R\over 2M^2}),
\label{19}\\
P_{\bar T} & = & 2M P_T .
\label{20}
\end{eqnarray}
Since the variable canonically conjugate to the 
new momentum, $\bar p$, is still the left mass,
 we have continued to denote it by $m$.

This completes the first step of the transformation to Kruskal variables.
In the second step,
we define the Kruskal variables through the following
 point transformation on the scaled variables:
\begin{equation}
(\bR -1) e^{\bR} = -UV, 
\label{23}
\end{equation}
\begin{equation}
{\bar T} = \ln |{V\over U}|.
\label{24}
\end{equation}
It follows that 
\begin{equation}
P_{\bar T} = {VP_V -UP_U \over 2},
\label{21}
\end{equation}
and
\begin{equation}
\PibR = {VP_V +UP_U \over 2F},
\label{22}
\end{equation}
where $F$ is a function of $UV$ through (\ref{5}) and (\ref{23}).
Thus, the new set of canonical variable is $(U,P_U, V, P_V)$ as well 
as the canonically conjugate parameters $(m, {\bar p})$.
This completes our presentation of the canonical transformation from
the curvature coordinate variables to the Kruskal variables.

Next, we present expressions for the constraints in terms of the 
new canonical variables.
It is easy to check that in terms of the Kruskal variables the diffeomorphism 
constraint is 
\begin{equation}
H_r = P_U U^{\prime} + P_V V^{\prime}.
\label{diffuv}
\end{equation}
It is easier to express  the rescaled scalar constraint, $\La H$, in terms
of the Kruskal variables, rather than $H$. To do this, we 
use (\ref{9}), (\ref{11}) and (\ref{17})-(\ref{24}) in (\ref{13}). Then
it follows that $\La H$ takes the form
 \begin{equation}
\La H =  P_V V^{\prime} -P_U U^{\prime} - {\bR^2\over 2M^2}e^{\bR} P_UP_V .
\label{scalaruv}
\end{equation}

Note that from (\ref{23}), the vanishing of $F$ implies the vanishing of 
at least one of $U$ or $V$. It follows from (\ref{24}) and (\ref{22}), that 
the transformation between curvature coordinate variables and Kruskal variables
is singular on the horizon. Note, however, that the expressions for the 
constraints in terms of the Kruskal variables remain non-singular on the 
horizon.

\section*{4. ADM variables in terms of Kruskal variables.}

 As noted in previous sections, both the 
transformation from the ADM variables to the curvature coordinate variables
as well the transformation from 
curvature coordinate variables to Kruskal variables, are singular 
when  $F=0$. In this section we present expressions for the 
ADM variables in terms of the Kruskal variables and see that this 
transformation is manifestly {\em non-singular} at $F=0$.

To express $\La$ in terms of the Kruskal variables we start from (\ref{14})
and use (\ref{5}),(\ref{9}) and (\ref{17})-(\ref{24}). 
Then it is straightforward
to show that 
\begin{equation}
\La^2 = {16M^2\over \bR e^{\bR}}
         (U^{\prime} + {P_V \bR^2 e^{\bR}\over 4M^2})
         (-V^{\prime} + {P_U \bR^2 e^{\bR}\over 4M^2}) . 
\label{25}
\end{equation}
In the above expression, $\bR$ and $M$ are to be thought of as functions
of the Kruskal variables. $\bR$ is a function of $UV$ through (\ref{23}).
To express $M$ in terms of the Kruskal variables, we use 
(\ref{9}),(\ref{20}),(\ref{21}) and (\ref{11b}). We obtain 
\begin{equation}
M^2 = m^2 + \int_{-\infty}^r d{\bar r} {UP_U - VP_V\over 2}. 
\label{MUV}
\end{equation}

To express $\La \PL$ in terms of the Kruskal variables we 
 start from (\ref{7}) and use (\ref{9}),(\ref{17}),(\ref{19}) and
(\ref{23}), (\ref{24}), (\ref{22}).  
We obtain
\begin{equation}
\La \PL = -4M^2 e^{-\bR}\big(V(U^{\prime} + {P_V \bR^2 e^{\bR}\over 4M^2})
                            +U(-V^{\prime} + {P_U \bR^2 e^{\bR}\over 4M^2})
                         \big) .
\label{26}
\end{equation}

Next, $R$ can be expressed as 
\begin{equation}
R= 2M\bR
\end{equation}
with $M$ given by (\ref{MUV}) and $\bR$ by (\ref{23}).

The calculation of $P_R$ in terms of the Kruskal variables is fairly involved
and we sketch the main steps here. We evaluate $P_R$ through (\ref{10}).
We first evaluate the expression 
$((\La \PL)^\prime (RR^{\prime})  
- \La \PL (RR^{\prime})^\prime))$  which occurs in (\ref{10}).
To this end, it is useful to define 
\begin{eqnarray}
g_1 &=& U(-V^{\prime} + {P_U \bR^2 e^{\bR}\over 4M^2}),
\label{defg1}\\
g_2&=& V(U^{\prime} + {P_V \bR^2 e^{\bR}\over 4M^2}).
\label{defg2}
\end{eqnarray}
Then from (\ref{9}),(\ref{20}),(\ref{17}) and (\ref{26}) it follows that 
\begin{eqnarray}
&(\La \PL)^\prime (RR^{\prime})  
- \La \PL (RR^{\prime})^\prime =\;\;\;\;\;\;\;\;\;\;\;\;\;\;\;\;\;\;\nonumber\\
 & 4M^2({R^\prime\over 2M})\big( -\La\PL\bR^\prime(\bR+1)
                                     -4M^2e^{-\bR}\bR
                                          (g_1^{\prime}+g_2^{\prime})
                                \big)
-4M^2 \bR ({R^\prime\over 2M})^\prime \La \PL .
\label{subpr}
\end{eqnarray}
The equations (\ref{17}),(\ref{20}),(\ref{9}),(\ref{23})and (\ref{21})
imply that 
\begin{equation}
{R^\prime\over 2M} = \bR^{-1}e^{-\bR}(g_1-g_2).
\label{rprime}
\end{equation}
Using  (\ref{rprime}) and (\ref{26}) in (\ref{subpr}) we get 
\begin{equation}
(\La \PL)^\prime (RR^{\prime})  
- \La \PL (RR^{\prime})^\prime =
32M^4 e^{-2\bR} (g_2 g_1^{\prime}- g_2^{\prime}g_1) .
\end{equation}
By substituting this expression in (\ref{10}) and using 
(\ref{5}), (\ref{22}) and (\ref{26}), we obtain 
\begin{eqnarray}
P_R &=& -\bR({VP_V +UP_U\over 4M}) - Me^{-\bR}(UV^{\prime} - VU^{\prime})
\nonumber\\
&& + {\La \PL \over 4M\bR} - 16 {M^3 e^{-\bR} \over \La^2}{\cal H}
\label{27}
\end{eqnarray}
where
\begin{equation}
{\cal H} = (U^{\prime} + {P_V \bR^2 e^{\bR}\over 4M^2})
         (-V^{\prime} + {P_U \bR^2 e^{\bR}\over 4M^2})^\prime  
- (U^{\prime} + {P_V \bR^2 e^{\bR}\over 4M^2})^\prime
         (-V^{\prime} + {P_U \bR^2 e^{\bR}\over 4M^2}) . 
\label{28}
\end{equation}
In (\ref{27}),  $\La\PL$ and $\La^2$ are given by (\ref{25}) and (\ref{26})
and $M$ by (\ref{MUV}) and $\bR$ through (\ref{23}).

As mentioned earlier, 
 (\ref{5}) and (\ref{23}) imply that  $F=0$ 
corresponds to the vanishing of at least one of $U$ or $V$. 
As can be explicitly verified, the expressions for 
the ADM variables $(\La, \PL, R, P_R)$ are all non-singular when this happens
and hence, at the horizon, the ADM variables continue to be smooth functions 
of the Kruskal variables.

For the transformation to be defined, it is necessary to impose the condition 
$M\neq 0$. Since we are interested in black holes rather than naked 
singularities, we shall impose $M>0$. Further, in the ADM description,
the conditions $\La >0$ and $R>0$ hold and these conditions must be imposed
in the description in terms of the Kruskal variables. We shall comment 
further on these points in section 6 where we invert the transformation 
and express the Kruskal variables in terms of the ADM variables.

Finally, note that we  have yet to reconstruct the remaining ADM 
variables, namely the parametrization times ($\tau_{\pm}$), from the 
Kruskal variables. Since this reconstruction requires not only the 
variable $\bar p$ in the Kruskal description but also the asymptotic
behaviour of the Kruskal variables, we shall return to this point in section
5, after we have analysed the asymptotics.

\section*{5. Asymptotics}

In this section we describe our choice of asymptotic conditions for the 
ADM variables and the Kruskal variables. These conditions ensure that the 
various integrals encountered in the canonical transformation of section 3
are convergent near spatial infinity.

In future work we would like our framwework to admit
matter couplings. To this end,
we expect that our choice of  boundary conditions is  
general enough to handle couplings to a large class of matter fields, 
namely those for 
 which the matter fields fall off faster than any power of 
 $|r|^{-1}$ at infinity. 
\footnote{Note that we have not looked for the weakest possible asymptotic 
conditions on the gravitational variables
 but simply for ones which provide an elegant, consistent description
and which admit coupling to a fairly large class of matter fields.}

In the Schwarzschild spacetime, the Kruskal coordinates are related to the 
curvature coordinates near right and left infinity by 
\begin{eqnarray}
U &=& \mp\sqrt{{R\over 2M}-1} \;\;e^{R-T\over 4M} , 
\label{51}\\
V &=&\pm\sqrt{{R\over 2M}-1} \;\;e^{R+T\over 4M} . 
\label{52}
\end{eqnarray}
Therefore,
as $r\rightarrow \pm\infty$ we impose
\begin{eqnarray}
U & = & \mp\sqrt{{|r|\over 2M_{\pm}}-1} \;\;e^{|r|\over 4M_{\pm}}
       \;\; e^{-{\bar T}({\pm}\infty)\over 2}( 1+ \Theta (r)) ,
\label{53}\\
V & = & \pm\sqrt{{|r|\over 2M_{\pm}}-1} \;\;e^{|r|\over 4M_{\pm}}
       \;\; e^{{\bar T}({\pm}\infty\over 2}( 1+ \Theta (r)) ,
\label{54}
\\
UP_U &= & \Theta (r),
\label{55}\\
VP_V &= & \Theta (r). \label{56}
\end{eqnarray}
Here 
${\bar T}({\pm}\infty):= {\bar T}(r)|_{r=\pm\infty} $
and $M_{\pm}= M(r)|_{r=\pm\infty}$, 
with $M(r)$ given by (\ref{MUV}).
$\Theta (r)$ denotes smooth fall off faster than $|r|^{-n}$,
$n$ arbitrarily large.

By virtue of the relation between Kruskal variables and ADM variables derived
in section 4, the conditions (\ref{53})-(\ref{56}) induce the following 
asymptotic behaviour for the ADM variables
\footnote{Note that (\ref{59}) admits the Schwarzscild solution. In contrast
(41) of \cite{kuchar} admits the Schwarzschild solution only if $\epsilon <1$
in that equation, whereas (58) of \cite{kuchar}
does not admit the Schwarzschild solution at all!
}
 as $r\rightarrow \pm \infty$:
\begin{eqnarray}
\La &=& {1\over \sqrt{1- {2M_{\pm}\over |r|}}}
               + \Theta (r),
\label{59}\\
R &= & |r| + \Theta (r),
\label{510}\\
\PL &=& \Theta (r) ,
\label{511}\\
P_R &= &\Theta (r) .
\label{512}
\end{eqnarray}

Alternatively, we can start from (\ref{59})-(\ref{512}) and following the 
transformations described in section 3, we can induce boundary conditions
on the curvature coordinate variables, the scaled curvature coordinate 
variables and finally, the Kruskal variables. Following this procedure,
it can be checked that (\ref{53})-(\ref{56}) are obtained.


Next, we analyse the asymptotic behaviour of the constraints, the 
lapse function and the shift vector field. It can be checked that the 
constraints fall off faster than any inverse power of $|r|$ i.e.
\begin{equation}
H, H_r = \Theta (r) .
\end{equation}
The behaviour of the lapse and shift should be such  that 
the smeared scalar constraint, $\int_{-\infty}^{\infty}dr NH$ and the 
smeared diffeomorphism constraint, $\int_{-\infty}^{\infty}dr N^rH_r$ 
be well defined, differentiable functions on the phase space and that the
motions they generate
preserve the boundary conditions (\ref{59})-(\ref{512}). 

For the 
 shift vector field as $r\rightarrow \pm \infty$, we impose
\begin{equation}
N^r = \Theta (r),
\end{equation}
It can be checked that with this behaviour the smeared diffeomorphism 
constraint is a well defined, differentiable function on the phase space.

For the lapse function we impose
\begin{equation}
N= N_{\pm} + \Theta (r),
\end{equation}
where $N_{\pm}$ are constants. In the description in terms of the 
ADM variables, the smeared scalar constraint is  a differentiable 
function on phase space  only when $N_+$ and $N_-$ vanish.
For non-vanishing $N_+$ or $N_-$,
the boundary term $(N_+ M_+ - N_-M_-)$ has to be added to the smeared
scalar constraint to render the combination differentiable on the phase 
space. As discussed in \cite{kuchar}, parametrization of  the times at 
spatial infinity leads to the action (\ref{1}) in which the 
boundary term is  replaced by the term 
$({\dot \tau_+} M_+  -{\dot \tau_-} M_- )$. In the description
in terms of Kruskal variables, the smeared scalar constraint is differentiable
{\em without} the addition of any boundary term even for 
non-vanishing $N_+$ or $N_-$.
As in the case of the 
description in terms of the curvature coordinate variables, instead of the 
boundary terms of the ADM description there are a pair
of  canonically conjugate parameters $({\bar p}, m)$ (see section 2).

As mentioned earlier $m$ is the mass at left spatial infinity.
$\bar p$ can still be 
interpreted as the difference between the Killing time and 
the parametrization time at left infinity as we now show.
From (\ref{19}) we have
\begin{equation}
\int_{-\infty}^{\infty}{\PiR R + P_T T\over 2M^2} = 
                   {\bar T}(-\infty ) - {T(-\infty )\over 2m} .
\label{516}
\end{equation}
Then (\ref{16}) implies that 
\begin{equation}
{\bar p} = p + 2m ({\bar T}(-\infty ) - {T(-\infty )\over 2m}).
\label{517}
\end{equation}
From (\ref{11a})  we have that
\begin{equation}
p = T(-\infty ) -\tau_- ,
\label{518}
\end{equation}
which together with (\ref{517}) implies that
\begin{equation}
{\bar p} = 2m {\bar T}(-\infty ) -\tau_- .
\label{519}
\end{equation}
Since $2m {\bar T}(-\infty )$ is the Killing time at left spatial infinity,
$\bar p$ has the interpretation of the difference between the 
Killing time at left infinity and the parametrization time at left infinity.
\footnote{
Note that in general ${\bar p} \neq p$. However,  on the constraint 
surface $p={\bar p}$. Since the interpretation of  $p,{\bar p}$
comes from their interpretation on a solution, they have identical 
interpretations even though they define  different functions on the 
(unconstrained) phase space.
}
Note also that from (\ref{8})and (\ref{19})  we have
\begin{equation}
{\bar T}(\infty ) = {\tau_+ \over 2M_+} .
\label{520}
\end{equation}
 Now we can finally
 complete the reconstruction of section 4 of  the ADM parameters
$\tau_{\pm}$ from the Kruskal variables.
Using (\ref{519}) and (\ref{520})
we have 
\begin{eqnarray}
\tau_- &=& -({\bar p} - 4m{\bar T}(-\infty )),\label{tau-} \\
\tau_+ & =& 2M_+{\bar T}(\infty ) .
\label{tau+}
\end{eqnarray}
Here $M_+= M(r)|_{r=\infty}$ is obtained from the Kruskal variables through
(\ref{MUV}) and ${\bar T}(\pm\infty )$ 
are obtained from the asymptotic behaviour of the 
Kruskal variables from (\ref{53}) and (\ref{54}).

This completes our discussion of the asymptotics, as well as the 
reconstruction, of the ADM variables from the Kruskal variables.
In the next section we shall invert the expressions for the ADM variables
in terms of the Kruskal variables.

\section*{6. Kruskal variables in terms of ADM variables.}

The Kruskal variables can be expressed in terms of the ADM variables using 
the results of section 3 in conjunction with (\ref{8})- (\ref{11b}). 
However, the transformation described in section 3 as well as the 
transformation from ADM variables to curvature coordinate variables 
are both singular at the horizon. For the transformation between the 
ADM variables and the Kruskal variables to be non-singular and 
{\em invertible}, it is necessary to prove that the Kruskal variables 
can be constructed from the ADM variables through a non-singular 
transformation. In this section we provide the required proof.
We shall be brief and only describe the main steps.

In what follows, we shall assume that 
\begin{equation}
\La >0,\;\;\;\; M>0,\;\;\;\;R>0.
\label{positive}
\end{equation}

We define
\begin{eqnarray}
\La_1 &=& - V^{\prime} + {\bR^2\over 4M^2}e^{\bR}P_U ,\label{61}\\
\La_2 &=&  U^{\prime} + {\bR^2\over 4M^2}e^{\bR}P_V .\label{62}
\end{eqnarray}
Substitution of (\ref{61}), (\ref{62}) in (\ref{25}), (\ref{26}) yields
\begin{eqnarray}
\La^2 &=& {16M^2\over \bR e^{\bR}} \La_1\La_2 , \label{63}\\
\La \PL &=& -4 M^2 e^{-\bR}(U\La_1+ V\La_2 ). \label{64}
\end{eqnarray}
Note that (\ref{positive}) together with 
 the asymptotic behaviour of the Kruskal variables implies that 
\begin{equation}
\La_1 < 0, \;\;\;\;\; \La_2 < 0.
\label{64a}
\end{equation}

By using (\ref{5}), (\ref{17}), (\ref{18}), (\ref{22}), (\ref{61}), 
(\ref{62}) and (\ref{26}) in (\ref{27}), we get
\begin{equation}
{\cal H}(r) = -{\La^2e^{\bR}\over 16M^3}
       ( P_R + {F \PiR R (\bR +1) \over 2M} + {\La \PL F\over 4M}). 
\label{65}
\end{equation}
Equation (\ref{65}) is to be viewed as an expression for ${\cal H}$
in terms of the ADM variables. Thus in (\ref{65}), $M$ is given by (\ref{4}),
$F$ by (\ref{5}) and $\bR$ by (\ref{17}). 

From (\ref{28}),(\ref{61}) and (\ref{62}) we have
\begin{equation}
\La_2 \La_1^{\prime}-\La_1 \La_2^{\prime} = {\cal H} .
\label{66}
\end{equation}
At this stage it is useful to define 
\begin{equation}
{\cal G}(r) := {\La^2 \bR e^{\bR} \over 16 M^2} .
\label{67}
\end{equation}
From (\ref{63}) we get
\begin{equation}
\La_1\La_2 = {\cal G}, 
\label{68}
\end{equation}
where ${\cal G}$ is expressible as a function of the ADM variables through
(\ref{67}).

From (\ref{66}) and (\ref{68}) we get a first order ordinary differential 
equation for $\La_1\over \La_2$. The boundary conditions 
(\ref{53})-(\ref{56}) along with (\ref{520}) can be used in 
(\ref{61}) and (\ref{62}) to fix the integration constant in the solution 
of the differential
equation. The solution to the differential equation and 
 (\ref{68}) can be solved to obtain
\begin{eqnarray}
\La_1 &=&\sqrt{\cal G} \;\;e^{\tau^+\over 4M_+}\;\;
           e^{\int_{\infty}^rd{\bar r}({{\cal H}\over {\cal G}})},
\label{69}\\
\La_2 &=&\sqrt{\cal G} \;\;e^{-{\tau^+\over 4M_+}}\;\;
           e^{-\int_{\infty}^rd{\bar r}({{\cal H}\over {\cal G}})}.
\label{70}
\end{eqnarray}
Note that  the signs of $\La_1, \La_2$ are fixed by the conditions
(\ref{64a}).
Equations (\ref{69}) and (\ref{70}) express $\La_1, \La_2$
in terms of the ADM variables. 
We shall now solve for $P_U$ and $P_V$ in terms of $\La_1,\La_2$ and the 
ADM variables. From (\ref{9a}) and (\ref{11}), it can be shown that 
\begin{eqnarray}
UP_U&=& -2M \La^{-2}(\La H - H_r) (R^{\prime}- {\La \PL\over R}),
\label{71}\\
VP_V&=& 2M \La^{-2}(\La H + H_r) (R^{\prime}+ {\La \PL\over R}).
\label{72}
\end{eqnarray}
To evaluate $(R^{\prime}\pm {\La \PL\over R})$ we use 
(\ref{64}), (\ref{17}), (\ref{20}), (\ref{21}) and (\ref{23}). We obtain
\begin{eqnarray}
R^{\prime}+ {\La \PL\over R}&=& -{4M\over \bR e^{\bR}}V\La_2,
\label{73}\\
R^{\prime}- {\La \PL\over R}&=& {4M\over \bR e^{\bR}}U\La_1 .
\label{74}
\end{eqnarray}
We substitute (\ref{73}), (\ref{74}) in (\ref{71}), (\ref{72}) and use
(\ref{61}), (\ref{62}) to obtain
\begin{eqnarray}
P_U&=& -{\La H - H_r\over 2\La_2} , \label{75}\\
P_V&=& -{\La H +H_r\over 2\La_1} . \label{76}
\end{eqnarray}
Since $\La_1$ and $\La_2$ are given by (\ref{69}) and (\ref{70}),
equations (\ref{75}) and  (\ref{76}) express $P_U$ and $P_V$ in terms of the 
ADM variables.

We now show that $U$ and $V$ can also be determined in terms of the 
ADM data in a non-singular way.
From (\ref{73}) and (\ref{74}) we obtain,
\begin{eqnarray}
V&=& -(R^{\prime}+ {\La \PL\over R}){\bR e^{\bR}\over 4M\La_2},
\label{77}\\
U&=&(R^{\prime}- {\La \PL\over R}) {\bR e^{\bR}\over 4M\La_1}.
\label{78}
\end{eqnarray}
Since $\La_1$ and  $\La_2$ have been expressed in terms of the ADM
variables, equations (\ref{77}) and (\ref{78}) express $U$ and 
$V$ in terms of the ADM variables in a manifestly non-singular form. 

Finally, it is easy to see that $(m, {\bar p})$ are also determined in terms
of the ADM variables. $m$ is trivially obtained from the asymptotic behaviour 
of $\La$ at left infinity. $\bar p$ can be expressed as 
\begin{equation}
{\bar p} = {-2mV(r)\over U(r)}|_{r=-\infty} - \tau_-  .
\end{equation}
Since $V$ and $U$ are known in terms of the ADM variables, so is $\bar p$.

Thus, we have shown that the Kruskal variables,
$(U, P_U, V, P_V, m, {\bar p})$, are uniquely determined through  manifestly 
non-singular  transformations of the ADM variables, 
$(\La, \PL, R, \PiR,\tau_+, \tau_-)$.

\section*{7. Concluding remarks.}

In this work we have constructed a
 transformation between the  ADM variables,\\  
\noindent $(\La, \PL, R, \PiR,\tau_+, \tau_-)$,  and 
 new  canonical variables  $(U, P_U, V, P_V, m, {\bar p})$. 
$U$ and $V$ are interpreted as Kruskal coordinates and $P_U, P_V$ 
are their conjugate momenta. $m$ is the mass at left infinity.
$\bar p$ is interpreted as the difference between the  Killing time at left 
infinity and the parametrization time, $\tau_-$, at left infinity, with 
the Killing time at right infinity synchronised with the parametrization 
time, $\tau_+$, at right infinity.

This transformation is manifestly non-singular and invertible.
In particular, it is non-singular at the horizon. For the transformation to 
be well defined, we assume that the mass function, $M(r)$, and the 
areal radius, $R(r)$, are both strictly positive i.e. $R, \;M >0$.
The interpretation of $\La^2$ as a metric coefficient implies that 
$\La^2 \neq 0$. The conditions $\La >0$ and $M >0$ lead to complicated 
restrictions on the Kruskal variables through 
(\ref{MUV}) and (\ref{25}). The condition $R>0$ is equivalent, through
(\ref{23}), to the condition $(-UV) >1$. Note that this is exactly
the condition that defines the singularity free region of the extended
Schwarzschild spacetime.

In terms of the Kruskal variables, the constraints are given by 
(\ref{diffuv}) and (\ref{scalaruv}). Equations (\ref{75}) and 
(\ref{76}) imply that the vanishing of the constraints is equivalent to 
the vanishing of $P_U$ and $P_V$. This equivalence does not entail the 
involved arguments  near $F=0$ which were used in \cite{kuchar} to show that 
the imposition of the constraints  implied $\PiR= P_T=0$.
After the imposition of the constraints, the true degrees of freedom 
are $({\bar p}, m)$ and quantization of this reduced theory is trivial.
The condition $M(r) >0$ reduces to the condition $m>0$ on the 
constraint surface. It is useful to make a further point transformation
on the pair $({\bar p}, m)$ to obtain $(x= \ln m,\; p_x= m{\bar p})$. 
We can now pass to quantum theory on the 
Hilbert space 
$\{\psi (x)\in {\cal L}^2({\bf R})\}$ by  setting 
\begin{eqnarray}
{\hat x}\psi (x)&= &x \psi (x) , \\
{\hat  {p_x}}\psi (x) &=&-i {d\over dx} \psi (x) .
\end{eqnarray}

As mentioned in footnote 1, a transformation to new canonical variables 
which also have the interpretation of Kruskal coordinates, was constructed
in \cite{kuchar} by rescaling the curvature coordinate variables by the 
left mass `$m$' rather than by the mass function, `$M(r)$', as is done here.
It can be checked that this transformation of \cite{kuchar} 
is singular on points in the 
(unconstrained) phase space when 
$F=1-{2M\over R}\neq 0$ and 
$f:=1-{2m\over R}= 0$. This is, of course, possible only off the constraint
surface. Nevertheless, it is clear  that any neighbourhood of the constraint
surface contains points where $F\neq 0, f= 0$ and hence, where the 
transformation is ill defined.  Since Poisson brackets involve functional
{\em derivatives}, their definition, even on the constraint surface, 
requires a nonsingular structure in 
a neighbourhood of the constraint surface. This is one reason why 
the transformation to Kruskal variables in \cite{kuchar} is unsatisfactory.
In contrast the transformation described in this work is nonsingular
on the entire phase space (of course, subject to the conditions
$M, \La^2, R >0$).

Although this work is concerned with spherically symmetric vacuum gravity,
the physically interesting problem is that of spherical matter collapse
say, the collapse of a massless scalar field.
In the collapse situation, the coordinate $r$ ranges from $0$ to 
$\infty$ with $r=0$ being the fixed point under the action of the 
rotation isometry 
group.  As discussed in \cite{romano}, the mass function $M(r)$ can still be 
constructed from the ADM data. Moreover the apparent horizon is defined 
through $F=0$. The canonical transformation to $R,T$  variables can still be 
done but this transformation is singular on the apparent horizon.
Thus, it would be of interest to construct the analog of the Kruskal 
coordinates for the matter collapse case. Note that the treatment in this 
work would have to be modified to deal with the condition 
$M(r)|_{r=0} =0$. 

Apart from the Kruskal coordinates, there exist other global coordinates
for the extended Schwarzschild spacetime such as those described in 
\cite{strobl,poisson}. It would be of interest to reconstruct these  
from the ADM data. In particular, it would be of interest to try to use
these (putative) variables to construct a time variable which is a spacetime
scalar \cite{karel,romano} even in the presence of matter couplings.

\section*{Acknowledgments}

\noindent 
I gratefully acknowledge helpful discussions of
matter pertinent to this work with Karel Kucha{\v r}. I thank
 Jorma Louko for encouragement and for his incisive comments. 
 I owe special thanks to Joseph Romano for helping me formulate
the ideas in this work and work them out in the context 
of the CGHS model. Part of this work was done at the University of Utah
and supported by the NSF grant PHY9207225.

\end{document}